\documentclass[aps,prd,preprint,superscriptaddress]{revtex4-1}
\pdfoutput=1
\usepackage{graphicx}  
\usepackage{subfigure}
\usepackage{latexsym}
\usepackage{slashed}
\usepackage{dcolumn}   
\usepackage{bm}        
\usepackage{amsmath}
\usepackage{amssymb}   
\usepackage{longtable}
\usepackage[section]{placeins}
\usepackage{float}

\newcommand{\be}{\begin{equation}}
\newcommand{\ee}{\end{equation}}
\newcommand{\bea}{\begin{eqnarray}}
\newcommand{\eea}{\end{eqnarray}}
\newcommand{\bma}{\begin{matrix}}
\newcommand{\ema}{\end{matrix}}

\newcommand{\bml}{\begin{mathletters}}
\newcommand{\eml}{\end{mathletters}}
\newcommand{\bes}{\begin{subequations}}
\newcommand{\ees}{\end{subequations}}

\newcommand{\bi}{\begin{itemize}}
\newcommand{\ei}{\end{itemize}}

\newcommand{\uva}{\affiliation{Department of Physics, University of Virginia, Charlottesville, VA 22904-4714, USA}}
\newcommand{\hue}{\affiliation{Center for Theoretical and Computational Physics, Hue University College of Education, Hue, Vietnam}}

\begin{document}





\title{Constraining Inflationary Dark Matter in the Luminogenesis Model }

\author{Pham Q. Hung}
\email{pqh@virginia.edu}\uva \hue
\author{Kevin J. Ludwick}
\email{kludwick@virginia.edu} \uva






\begin{abstract}

Using renormalization-group flow and cosmological constraints on inflation models, we exploit a unique connection between cosmological inflation 
and the dynamical mass of dark matter particles in the luminogenesis model, a unification model with the gauge group $SU(3)_C \times SU(6) \times U(1)_Y$, which breaks to the 
Standard Model with an extra gauge group for dark matter 
when the inflaton rolls into the true vacuum.  In this model, 
inflaton decay gives rise to dark matter, which in turn decays to luminous matter in the right 
proportion that agrees with cosmological data.  Some attractive features of this model include self-interacting dark matter, which may resolve the problems of 
dwarf galaxy structures and dark matter cusps at the centers of galaxies. 

\end{abstract}
\maketitle

\renewcommand{\thepage}{\arabic{page}}
\setcounter{page}{1}
\renewcommand{\thefootnote}{\#\arabic{footnote}}


\begin{center}
{\bf Introduction}
\end{center}

Dark matter comprises a large portion of our universe, and yet it still eludes our full comprehension.  There are many unsolved mysteries about the nature 
of dark matter, such as the "missing satellite" problem and the dark matter cusps at the 
centers of galaxies found in simulations but not in observations.

What if all matter originally came from the dark sector?  One 
possible way to combine dark matter and the Standard Model is via the luminogenesis model \cite{Luminogenesis1, Luminogenesis2}.  It is a model that undergoes  
the symmetry breaking $SU(3)_C \times SU(6) \times U(1)_Y \rightarrow SU(3)_C \times SU(4)_{DM} \times SU(2)_L \times U(1)_Y  \times U(1)_{DM}$ at the DUT 
(Dark Unified Theory) scale.  This breaking occurs when the inflaton slips into the true vacuum of its potential.  
The inflaton decays to dark matter, while decay to luminous matter is suppressed at the tree level.  Below the luminogenesis scale 
$M_{lum}$, dark matter decays to radiation and mirror matter \cite{PQmirror1} (which will be discussed below), and almost all mirror fermions decay to standard fermions.  Freeze-out of these conversions occurs 
at high energy scales, leaving standard Big Bang Nucleosynthesis unaffected.  In this model, $SU(4)_{DM}$ is unbroken and is confining at a scale much smaller than the DUT scale. This is the main emphasis of this manuscript.

Dark matter in this model is self-interacting, which may resolve the discrepancies between observations and simulations mentioned 
above.  Another attractive feature of this model is the absence of proton decay, which may explain its lack of observation.   

In this paper, we explore the interesting and unique connection between cosmic inflation and dark matter.  We first analyze renormalization-group flow from the 
DUT scale to the confinement scale of $SU(4)_{DM}$, which provides the dynamical mass of the dark matter particles.  We obtain the $SU(4)_{DM}$ coupling at the DUT scale 
from the unification with the $SU(2)_L$ coupling at the DUT scale, which we get from running the $SU(2)_L$ coupling from the electroweak scale 
to that scale.  Then we examine constraints on the DUT scale based 
on constraints on inflation from Planck for a symmetry-breaking (Coleman-Weinberg) inflation potential and obtain constraints on the dark matter dynamical mass.  

Before embarking on the main topic of this manuscript, namely the connection between cosmic inflation and dark matter, it is important, for the sake of clarity, to give a short summary of the luminogenesis model \cite{Luminogenesis1, Luminogenesis2} and of the electroweak-scale right-handed neutrino model (EW$\nu_R$ model) \cite{PQmirror1} where mirror fermions appear.
\bigskip
\begin{center}
{\bf Mini-Review of the Luminogenesis Model and the EW$\nu_R$ Model}
\end{center}
The model proposed by \cite{Luminogenesis1, Luminogenesis2} which incorporates the EW$\nu_R$ model \cite{PQmirror1,PQmirror2} can be summarized as follows.

The idea in \cite{Luminogenesis1, Luminogenesis2} was to unify the dark matter and luminous sectors into a gauge group, $SU(6)$,  which contains the Standard Model (SM) $SU(2)_L$ under which dark matter particles are singlets and $SU(4)$, the DM gauge group which confines at some scale $\Lambda_4$. In this model, $SU(3)_C$ and $U(1)_Y$ are just "spectator" gauge groups. Specifically, we have (with the acronym DUT referring to that type of unification)

\be
\label{DUT1}
SU(3)_C \times SU(6) \times U(1)_Y \stackrel {M_{DUT}} {\longrightarrow} SU(3)_C \times SU(4)_{DM} \times SU(2)_L \times U(1)_{DM} \times U(1)_Y
\ee
\be
\label{DUT2}
SU(3)_C \times SU(4)_{DM} \times SU(2)_L \times U(1)_{DM} \times U(1)_Y \stackrel {\Lambda_{ DM}} {\longrightarrow} SU(3)_C \times SU(4)_{DM} \times SU(2)_L \times U(1)_Y
\ee
\be
\label{DUT3}
SU(3)_C \times SU(4)_{DM} \times SU(2)_L \times U(1)_Y \stackrel {\Lambda_{EW}} {\longrightarrow} SU(3)_C \times SU(4)_{DM} \times U(1)_{em}.
\ee
As one can see from the above symmetry breaking pattern, the Luminogenesis model contains two unbroken non-abelian gauge groups: $SU(3)_C $ and $SU(4)_{DM}$, both of which are assumed to confine at $\Lambda_{QCD}$ and $\Lambda_{4}$ respectively. The aim of this paper is to determine what $\Lambda_{4}$ (replaced by the symbol $\mu_{DM}$ for convenience below) might be since this scale is related to the dynamical mass of dark matter. To complete this mini-review, we will list the fermion and scalar representations and summarize their roles in the Luminogenesis model.

In \cite{Luminogenesis1, Luminogenesis2}, a set of criteria was written down to guide the choice of fermion representations.

1) SM particles are required to be singlets under the DM gauge group $SU(4)_{DM}$. In particular, left-handed SM particles are required to be doublets under $SU(2)_L$. From Table~\ref{table1}, one can see that the $\underline{6}$ representation of $SU(6)$ satisfies this criterion since it contains ${\bf (1,2)}$ under $SU(4) \times SU(2)$. A representation under $SU(3)_C \times SU(6) \times U(1)_Y$ which contains the SM left-handed quark doublets and right-handed quark singlets is then represented by ${\bf (3,6, 1/6)_L + (3,1,2/3)_R}$ with the last entries being the $U(1)_Y$ quantum number. Similarly, the SM leptons are contained in ${\bf (1,6, -1/2)_L + (1,1,-1)_R}$.

2) Since the gauge group at the DUT is $SU(3)_C \times SU(6) \times U(1)_Y$ and $\underline{6}$ is a complex representation, anomaly freedom requires that for every left-handed fermion multiplet there exists an equivalent right-handed multiplet so that the anomaly cancels between left and right. The right-handed $SU(2)$ doublets and left-handed singlets are called "mirror fermions" in \cite{PQmirror1}. They are contained in ${\bf (3,6, 1/6)_R + (1,6, -1/2)_R + (3,1,2/3)_L  + (1,1,-1)_L}$. The introduction of the mirror fermions in  \cite{PQmirror1} was motivated by the possibilities of having Majorana masses for right-handed neutrinos of the order of the electroweak scale, i.e. $O(\Lambda_{EW} \sim 246 \, \mathrm{GeV})$ naturally and of the possible detection of these right-handed neutrinos directly at the Large Hadron Collider (LHC). This would yield a direct test of the seesaw mechanism which provides an elegant explanation for the smallness of neutrino masses.  Details of the EW$\nu_R$ model can be found in  \cite{PQmirror1,PQmirror2}.

As discussed in \cite{Luminogenesis1, Luminogenesis2}, by embedding the SM gauge group $SU(2)$ into $SU(6)$, the existence of these mirror fermions comes out naturally from the requirement of anomaly freedom of $SU(3)_C \times SU(6) \times U(1)_Y$. Extensive studies of the contributions of mirror fermions to electroweak precision parameters have been carried out in \cite{PQmirror2} with the conclusion that the EW$\nu_R$ model fits the data very well despite the presence of right-handed mirror fermions. 

Mirror quarks and leptons can be searched for at the LHC \cite{UVAOKS}. The search for the electroweak-scale right-handed neutrinos is particularly interesting since it will be a direct test of the seesaw mechanism. As emphasized in \cite{PQmirror1}, the production of $\nu_R \nu_R$ at the LHC can give rise to interesting signals such as like-sign dileptons. Furthermore, the lightest mirror quark can decay into a SM quark by emitting a SM-singlet Higgs scalar ($q^M_R \rightarrow q_L + \phi_S$) through an interaction Lagrangian of the form: $g_{Sq} \,\bar{q}_L \ \phi_S \ q_R^M + h.c.$ where $q_L$ and $q_R^M$ refer to a SM left-handed and mirror right-handed quark doublet respectively.  A similar decay process applies to the lightest mirror lepton ($l^M_R \rightarrow l_L + \phi_S$) with $g_{Sl} \,\bar{l}_L \ \phi_S \ l_R^M + h.c.$.

Why would we be interested in the decays of mirror fermions into SM fermions? First, this would be an important process by which we can detect mirror fermions at colliders such as the LHC. Second, as expounded in \cite{Luminogenesis2}, dark matter (denoted by $\chi$) is contained in the ${\bf 20}$ and the inflation is the singlet part of ${\bf 35}$ (see Table~\ref{table1}). Since ${\bf 20 \times 20 = 1_s+35_a+175_s+189_a}$, the inflation decays mainly into dark matter through the interaction $g_{20} \, \Psi_{20}^{T} \sigma_2 \Psi_{20} \, \phi_{35}$ which contains $g_{20} \, \chi_{L}^{T} \sigma_2 \chi^{c}_{L} \phi_{inf}$. (This is one of the main points of \cite{Luminogenesis2}: the predominance of dark over luminous matter.) The conversion of a fraction of dark matter to luminous matter is achieved through the interactions with two scalar fields: $\Phi_{15}^{(L)}$ and 
$\Phi_{\bar{15}}^{(R)}$, and given by $\frac{g_{6}^2}{M_{15}^2} \, (\chi^{T}_{L} \sigma_2 l_L)\, (\chi^{c,T}_{L} \sigma_2 l^{M,c}_{L}) + h.c.$, resulting in $\chi_L + \chi_R \rightarrow l_L + l^{M}_R $ and $\bar{\chi}_L + \bar{\chi}_R \rightarrow \bar{l}_L + \bar{l}^{M}_R$. Although the decay length of the process $l^M_R \rightarrow l_L + \phi_S$ could be macroscopic at the LHC ("long-lived" $l^M_R$), in the early universe, $l^M_R$ basically decays promptly into SM leptons.  This, in a nutshell, is what \cite{Luminogenesis2} refers to as "luminogenesis." How effective the conversion of dark matter into SM particles is a question that depends on the prefactor $\frac{g_{6}^2}{M_{15}^2}$.

As is noted in \cite{Luminogenesis2}, a small dark matter asymmetry $\Delta n_\chi=n_\chi - n_{\bar{\chi}}$ is assumed to be present, with $n_\chi = n_{sym} + \Delta n_\chi$ only slightly bigger than $n_{\bar{\chi}}= n_{sym}$, where $n_{sym}$ is the symmetric part of $n_{\chi}$ and $n_{\bar{\chi}}$.  The asymmetric part consists of the small excess $\Delta n_\chi \ll n_{sym}$.  We assume that there is a global $U(1)_{\chi}$ symmetry for dark matter. The interactions involving the gauge bosons of the coset group $SU(6)/SU(4) \times SU(2) \times U(1)_{DM}$ explicitly break the $U(1)_{\chi}$ symmetry, and their decays involving the interferences between the tree-level and one-loop diagrams will ultimately produce a net DM number, assuming the presence of CP violation in the DM sector.  (This process is similar to the one involving X and Y gauge bosons in $SU(5)$ GUT Theory.)  $\chi$ and $\bar{\chi}$ can annihilate via $\gamma_{DM}$, the massive dark photon of $U(1)_{DM}$, into luminous particle-antiparticle pairs via the effective interaction $\frac{g^2}{M_{\gamma_{DM}}}(\bar{\chi}  \gamma_{\mu} \chi)(\bar{f} \gamma^{\mu} f)$, and the particle-antiparticle pairs of luminous fermions annihilate to radiation.  The coefficients for this annihilation process and the conversion process involving the aforementioned scalars $\Phi_{15}^{(L)}$ and $\Phi_{\bar{15}}^{(R)}$ are independent of each other, and they are such that 14\% of all dark matter (14\% of asymmetric and symmetric parts) converts via the scalars and 86\% of asymmetric and symmetric parts annihilates via $\gamma_{DM}$ ultimately to radiation.  The 14\% of the symmetric parts of $n_\chi$ and $n_{\bar{\chi}}$ that is converted to luminous matter via the scalars has equal parts of luminous particles and anti-particles, and these annihilate to radiation.  So the whole of the symmetric parts of $n_\chi$ and $n_{\bar{\chi}}$ is mainly converted to radiation.  But since annihilation via $\gamma_{DM}$ requires the presence of both $\chi$ and $\bar{\chi}$, this annihilation does not affect the asymmetric part $\Delta n_\chi$, so overall, we are left with radiation from the symmetric parts of $n_\chi$ and $n_{\bar{\chi}}$ and 14\% luminous matter and 86\% DM in the asymmetric part, giving the correct proportion of luminous to dark matter as the conversion process via the scalars freezes out.  So we see that this process of luminogenesis depends upon both freeze-out and asymmetry in dark matter (and the asymmetry is propagated through to luminous matter), and the coefficients of the processes discussed are such that luminogenesis gives what is observationally expected.  These processes involved in luminogenesis are discussed in more detail in \cite{Luminogenesis2} and are being explored further in \cite{hunglumino}.

Notice that, in principle, the ${\bf 15}$ and ${\bf \bar{15}}$ scalars could mix with the scalar ${\bf 35}$ (which contains the inflaton) of the form $\lambda_{15,35} {\bf 15 \times \bar{15}\times 35 \times \bar{35}}$. We assume that $\lambda_{15,35}$ is small enough that such a mixing does not affect the inflationary potential and other physical processes discussed in \cite{Luminogenesis2}. It is however interesting to study consequences, if any, of such a mixing but this is beyond the scope of this paper.

\bigskip
\begin{center}

{\bf Predictions for the Dark Matter Dynamical Mass from RG Flow}
\end{center}

Combining renormalization-group (RG) flow with constraints on cosmic inflation from Planck \cite{Planck, Plancki}, 
we can make a prediction for the dark matter dynamical mass.  As discussed in \cite{Luminogenesis1, Luminogenesis2}, 
$SU(6) \rightarrow SU(4)_{DM} \times SU(2)_{L} \times U(1)_{DM}$.  The $\beta$-function equation for $SU(N)$, ignoring the negligible
term for the contribution due to scalar fields, is 
\begin{equation}
\frac{d\alpha}{d\ln \mu} \equiv \beta(\alpha) = - \left[ \frac{11}{3} N - \frac{2}{3} C(R)  n_{f_L} - \frac{2}{3} C(R)  n_{f_R} \right] \frac{\alpha^2}{2 \pi} + O(\alpha^3),
\label{beta}
\end{equation}
where the second and third terms correspond to the contributions from left-handed and right-handed fermions respectively.  

The representations and group structure of the luminogenesis model for each of the three families are given below \cite{Luminogenesis2}.  In passing, the existence of 
mirror fermions, as proposed by \cite{PQmirror1}, provides a mechanism in which right-handed neutrinos obtain Majorana masses proportional to the electroweak scale, and they 
could be searched for at the Large Hadron Collider.  
 \begin{table}[h]
 \begin{center}
 \begin{tabular}{|l|l|lr|||} \hline
 $SU(6)$ & $SU(4)_{DM} \times SU(2)_L \times U(1)_{DM}$ \\ \hline
 ${\bf 6}$ & ${\bf (1,2)_2 + (4,1)_{-1}}$  \\
 $ {\bf 20}$ & ${\bf (4,1)_3 + (4^\ast , 1)_{-3} + (6,2)_{0} }$ \\
 ${\bf 35}$ & ${\bf (1,1)_0 + (15,1)_0 + (1,3)_0 +(4,2)_{-3}}$  \\ 
 & ${\bf  + (4^\ast , 2)_3}$ \\  \hline
 \end{tabular}
 \end{center}
 \caption{\label{table1} ${\bf (1,2)_2}$ represents luminous matter while ${\bf (4,1)_3 + (4^\ast , 1)_{-3}}$ represent dark matter.}
 \end{table}
 \begin{table}[h]
 \begin{center}
 \begin{tabular}{|l|l|lr|||} \hline
  & $SU(3)_c \times SU(6) \times U(1)_{Y}$ \\ \hline
 R $\supset$ SM fermions & ${\bf (3,6, 1/6)_L + (1,6, -1/2)_L }$  \\
 &${\bf + (3,1,2/3)_R + (3,1,-1/3)_R }$ \\ 
 &${\bf + (1,1,-1)_R}$ \\
 \hline
 R $\supset$ mirror fermions &  ${\bf (3,6, 1/6)_R + (1,6, -1/2)_R }$  \\
 &${\bf + (3,1,2/3)_L + (3,1,-1/3)_L }$ \\ 
 &${\bf + (1,1,-1)_L}$ \\
 \hline
 R $\supset$ dark matter fermions & ${\bf (1,20,0)}$ \\  \hline
 \end{tabular}
 \end{center}
 \caption{\label{table2} R in the left column denotes representation. Standard Model (SM) left-handed doublets and right-handed singlets comprise the first row, 
 mirror right-handed doublets \cite{PQmirror1} and left-handed singlets comprise the second row, and dark matter left- and right-handed fermions belong to the last row.}
 \end{table}

The inflaton is represented by ${\bf (1,1)_0}$ of ${\bf 35}$.  It is assumed that ${\bf (15,1)_0 + (1,3)_0}$ ${\bf +(4,2)_{-3}+(4^{\ast},2)_3}$ of ${\bf 35}$ 
have masses that are on the order of the DUT scale and are therefore not included in our analysis of RG flow.  
  The inflaton decays to dark matter through a coupling via 
${\bf 20 \times 20 = 1_s+35_a+175_s+189_a}$, 
while decay to luminous matter is suppressed at the tree level.  Dark matter can decay to luminous matter through a coupling via ${\bf 20 \times \bar{6} = 15+105}$ and 
${\bf 20 \times 6 = \bar{15} + \bar{105}}$ and through the massive gauge boson of $U(1)_{DM}$, the dark photon.  More details are in \cite{Luminogenesis2}.

The $SU(4)_{DM}$ dark matter fermions are represented by ${\bf (4,1)_3 + (4^\ast , 1)_{-3}}$ in the ${\bf 20}$ representation of $SU(6)$.  
The ${\bf (6,2)_0}$ part of ${\bf 20}$ is assumed to decouple below its mass scale $M_2$.  Since dark matter should have no $U(1)_Y$ charge, 
the $SU(4)_{DM}$ particles in ${\bf (4,1)_{-1}}$ in the ${\bf 6}$ representation of $SU(6)$ cannot be dark matter, and they are assumed to decouple below the mass scale $M_1$.  

For $SU(2)_L$, the Casimir factor for the representation $R$ for all the right-handed and left-handed contributions is $C({\bf 2}) = \frac{1}{2}$.  The numbers of right- and 
left-handed fermions are $n_{f_R}=n_{f_L} = (3+1) 3=12$ for scales $\mu < M_2$ because, using Tables \ref{table1} and \ref{table2}, 
we see that ${\bf (3,6,1/6)}_{L,R}$ is ${\bf (3,1,2,1/6)}_{L,R}+{\bf (3,4,1,1/6)}_{L,R}$ under $SU(3) \times SU(4)_{DM} \times SU(2)_L \times U(1)_Y$
 and it contains three $SU(2)_L$ doublets (due to color) per family, and ${\bf (1,6,-1/2)}_{L,R}$, which is ${\bf (1,1,2,-1/2)}_{L,R} + {\bf (1,4,1,-1/2)}_{L,R}$ 
 under $SU(3) \times SU(4)_{DM} \times SU(2)_L \times U(1)_Y$, 
 contains one $SU(2)_L$ doublet per family.  When $\mu>M_2$, we have an 
additional (left-handed) contribution to the number of fermions of $6 \cdot 3=18$ from the six $SU(2)_L$ doublets per family from ${\bf (6,2)_0}$.  
Therefore, following Equation (\ref{beta}), the $SU(2)_L$ one-loop $\beta$-function is $\beta_2(\alpha_2) = -f_2 \frac{\alpha_2^2}{2 \pi}$, where 
\begin{eqnarray}
&f_2^{\mu<M_2}= (\frac{11}{3} \cdot 2 - \frac{2}{3} \cdot \frac{1}{2} \cdot 12 - \frac{2}{3} \cdot \frac{1}{2} \cdot 12)=-\frac{2}{3}, \nonumber \\
&f_2^{\mu>M_2} = (\frac{11}{3} \cdot 2 - \frac{2}{3} \cdot \frac{1}{2} \cdot 12 - \frac{2}{3} \cdot \frac{1}{2} \cdot 12 - \frac{2}{3} \cdot \frac{1}{2} \cdot 18)=-\frac{20}{3}.  
\label{f2}
\end{eqnarray}

For $SU(4)_{DM}$, $C({\bf 4}) = \frac{1}{2}$ and $C({\bf 6})=1$.  Below the scales $M_1$ and $M_2$, the numbers of right- and left-handed fermions are $n_{f_R} = n_{f_L} = 3$ 
 because, per family, ${\bf (4,1)_3}$ and ${\bf (4^{\ast},1)_{-3}}$ contribute one left-handed and one right-handed $SU(4)_{DM}$ fermion.  When the $M_1$ scale is relevant, 
 ${\bf (4,1)_{-1}}$ of ${\bf 6}$ of $SU(6)$ contributes $(3+1) 3=12$ to $n_{f_L}$ and $n_{f_R}$ since, per family, ${\bf (3,1,2,1/6)}_{L,R}+{\bf (3,4,1,1/6)}_{L,R}$ 
 contributes $3$ (due to color) and 
 ${\bf (1,1,2,-1/2)}_{L,R} + {\bf (1,4,1,-1/2)}_{L,R}$ contributes $1$.  When $M_2$ is relevant, ${\bf (6,2)_0}$ contributes $2$ (due to the $SU(2)_L$ doublet) per family to the number of left-handed fermions.  
 The $\beta$-function for $SU(4)_{DM}$ is therefore $\beta_4(\alpha_4) = - f_4 \frac{\alpha_4^2}{2 \pi}$, where 
 \begin{eqnarray}
&f_4^{\mu<M_1,M_2}= (\frac{11}{3} \cdot 4 - \frac{2}{3} \cdot \frac{1}{2} \cdot 3 - \frac{2}{3} \cdot \frac{1}{2} \cdot 3)=\frac{38}{3}, \nonumber \\
 &f_4^{M_1<\mu<M_2}=( \frac{11}{3} \cdot 4 - \frac{2}{3} \cdot \frac{1}{2} \cdot 3 - \frac{2}{3} \cdot \frac{1}{2} \cdot 3 - \frac{2}{3} \cdot \frac{1}{2} 
 \cdot 12-\frac{2}{3} \cdot \frac{1}{2} \cdot 12)=\frac{14}{3} ~({\rm when}~ M_1<M_2), \nonumber \\ 
&f_4^{M_2<\mu<M_1}=( \frac{11}{3} \cdot 4 - \frac{2}{3} \cdot \frac{1}{2} \cdot 3 - \frac{2}{3} \cdot \frac{1}{2} \cdot 3 - \frac{2}{3} \cdot 1 \cdot 6)=\frac{26}{3} ~({\rm when}~ 
 M_2<M_1), \nonumber \\ 
 &f_4^{\mu>M_1,M_2}=( \frac{11}{3} \cdot 4 - \frac{2}{3} \cdot \frac{1}{2} \cdot 3 - \frac{2}{3} \cdot \frac{1}{2} \cdot 3 - \frac{2}{3} \cdot \frac{1}{2} 
 \cdot 12-\frac{2}{3} \cdot \frac{1}{2} \cdot 12 - \frac{2}{3} \cdot 1 \cdot 6) =\frac{2}{3}.
\label{f4}
\end{eqnarray}

Solving the $\beta$-function equation $\beta(\alpha) \equiv \frac{d\alpha}{d\ln \mu} = -f \frac{\alpha^2}{2 \pi}$ gives us 
 \begin{equation}
\alpha(\mu_1)^{-1} = \alpha(\mu_2)^{-1} + \frac{f}{2 \pi} \ln \frac{\mu_1}{\mu_2}.
\label{alpha}
\end{equation} 
Using Equation (\ref{alpha}) for $SU(2)_L$ for different energy ranges and evaluating at $\mu_{DUT}$, we get 
\begin{equation}
\alpha_2(\mu_{DUT})^{-1} = \alpha_2(\mu_{EW})^{-1} + \frac{f_2^{\mu<M_2}}{2 \pi} \ln \frac{M_2}{\mu_{EW}} + \frac{f_2^{\mu>M_2}}{2 \pi} \ln \frac{\mu_{DUT}}{M_2},
\label{alpha2}
\end{equation}
where $\mu_{EW} = 246 ~{\rm GeV}$ is the electroweak energy scale, and $\alpha_2(\mu_{EW})  \approx 0.03$.  
Using Equation (\ref{alpha}) for $SU(4)_{DM}$ for different energy ranges and evaluating at $\mu_{DUT}$ gives us 
\begin{equation}
\alpha_4(\mu_{DUT})^{-1} = \alpha_4(\mu_{DM})^{-1} + \frac{f_4^{\mu<M_<}}{2 \pi} \ln \frac{M_<}{\mu_{DM}} + \frac{f_4^{M_< < \mu < M_>}}{2\pi} \ln \frac{M_>}{M_<} 
+ \frac{f_4^{\mu>M_>}}{2\pi} \ln \frac{\mu_{DUT}}{M_>},
\label{alpha4}
\end{equation}
where $\mu_{DM}$ is the dark matter dynamical mass and $M_<$ ($M_>$) is the lesser (greater) of $M_1$ and $M_2$.  
We assume that $M_1$ and $M_2$ are bigger than $\mu_{DM}$ 
and the electroweak scale or any observable scale.  If $M_1$ or $M_2$ is on the order of $\mu_{DUT}$ (and therefore 
not affecting RG flow), one may set it 
equal to $\mu_{DUT}$ in Equations (\ref{alpha2}) and (\ref{alpha4}) to get the appropriate expression.  Using Equations (\ref{alpha2}) and (\ref{alpha4}) and the unification 
of $SU(2)_L$ and $SU(4)_{DM}$ at the DUT scale ($\alpha_4(\mu_{DUT})=\alpha_2(\mu_{DUT})$), we get an expression for the dark matter dynamical mass:
\begin{equation}
\begin{split}
&\mu_{DM} = M_< ~{\rm Exp} \left[ -\frac{2 \pi}{f_4^{\mu<M_<}}(\alpha_2(\mu_{EW})^{-1} - \alpha_4(\mu_{DM})^{-1}) \right] \\
&\left( M_2^{f_2^{\mu>M_2}-f_2^{\mu<M_2}} \mu_{EW}^{f_2^{\mu<M_2}} M_<^{- f_4^{M_< <\mu < M_>}} M_>^{f_4^{M_< < \mu <M_>} - f_4^{\mu>M_>}} \mu_{DUT}^{f_4^{\mu>M_>} - f_2^{\mu>M_2}} \right)^{1/f_4^{\mu<M_<}}.
\end{split}
\label{muDM}
\end{equation}

Essentially, we run the $SU(2)_L$ gauge coupling from the known electroweak scale up to the DUT scale, which can be observationally constrained, and then 
we run the $SU(4)_{DM}$ gauge coupling down to its appropriate scale for dark matter.  And the dark matter's dynamical mass should be approximately equal to the energy scale of confinement for $SU(4)_{DM}$, just 
as the major contribution to the quarks' masses comes mainly from the $SU(3)_C$ confinement scale in the Standard Model.  In Figure \ref{RGflow2fig}, we show regions of 
confinement energy from $0.5 \leq \alpha_4 \leq 1.5$ for the one-loop 
RG flow for various values of $\mu_{DUT}$, and we see that the variation in energy scale in these regions is not very significant.  So we estimate that 
$\alpha_4(\mu_{DM}) \sim 1$.  

In preliminary analysis, we let $M_2$ vary from $10^5$ GeV to $\mu_{DUT}$, and we found that the dark matter dynamical mass was in general very high, ranging even 
up to $10^{13}$ GeV.  Because we are not interested in such purely academic values of the dark matter dynamical mass, we assume $M_2 \sim \mu_{DUT}$ in all our analysis 
that follows.

We emphasize the unique and interesting connection between 
the DUT scale and the dynamical mass of dark matter we have presented.  This connection has been made independently of the model of inflation.  
In what follows, we specify an inflation model in 
order to apply inflation constraints from cosmological probes to $\mu_{DUT}$.

\begin{figure}
\begin{center}
\fbox{\includegraphics[scale=1.6]{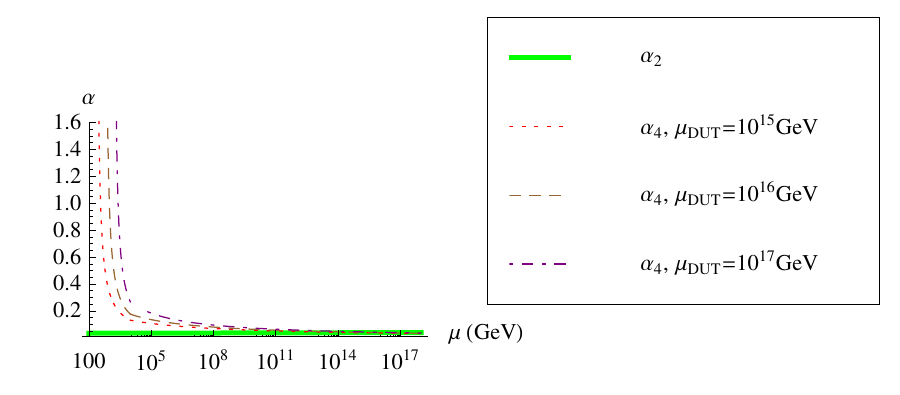}}
\caption{The unification of $SU(4)_{DM}$ and $SU(2)_L$ at the DUT scale for different values of $\mu_{DUT}$.  For these examples, we have chosen 
$M_1=10^{4} ~{\rm GeV}$ and $M_2 \sim \mu_{DUT}$.}
\label{RGflowfig}
\end{center}
\end{figure}

\begin{figure}
\begin{center}
\fbox{\includegraphics[scale=1.6]{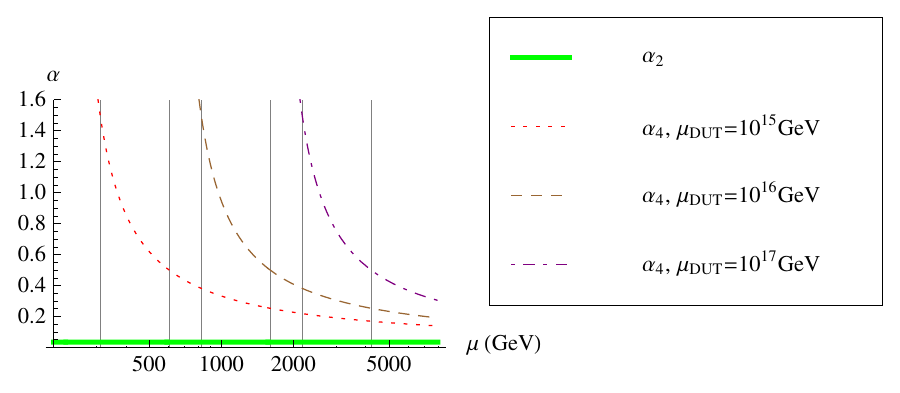}}
\caption{The regions of confinement for $0.5 \leq \alpha_4(\mu_{DM}) \leq 1.5$ for the functions from Figure \ref{RGflowfig} with the same values for $M_1$ and 
$M_2$.  The variation in 
energy scale in these regions is not very significant.  We use $\alpha_4(\mu_{DM}) \sim 1$.}
\label{RGflow2fig}
\end{center}
\end{figure}

\bigskip

\begin{center}
{\bf Constraints on Inflation from} {\bf Planck}
\end{center}

\begin{figure}
\begin{center}
\fbox{\includegraphics[scale=0.9]{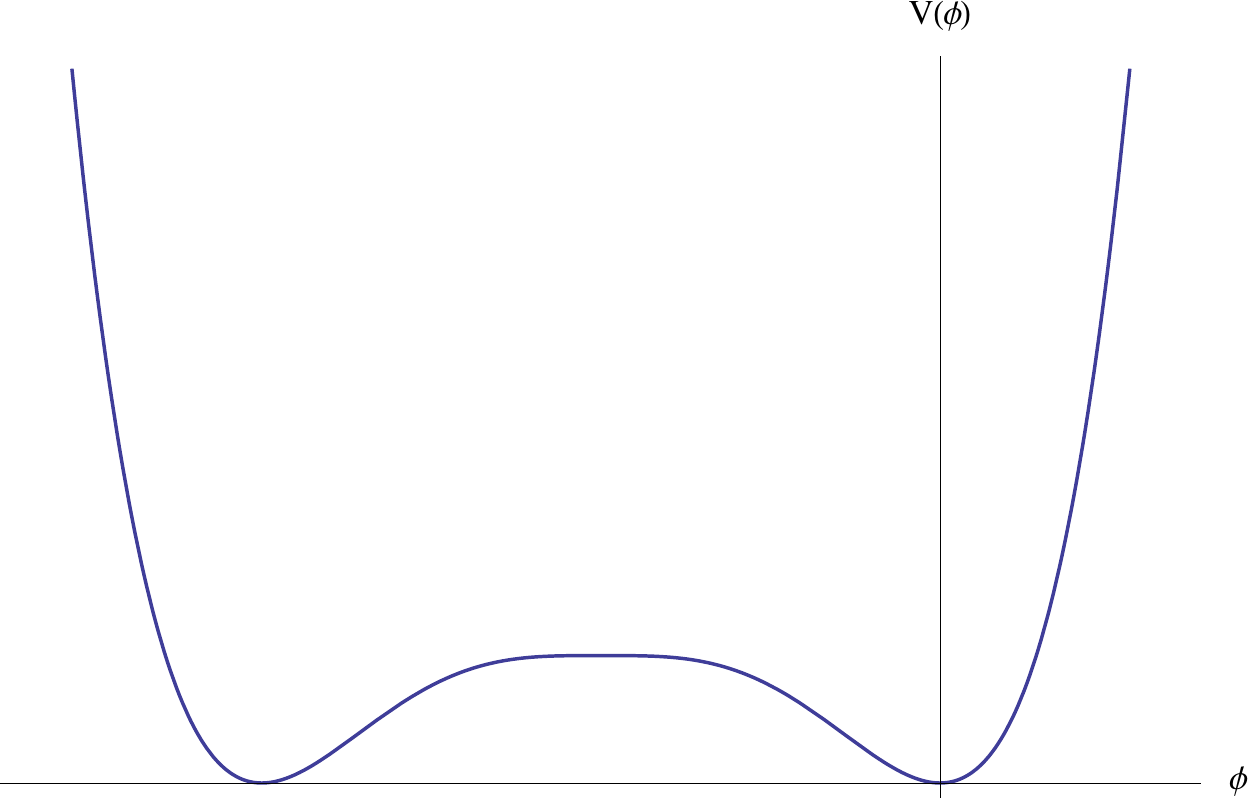}}
\caption{Our Coleman-Weinberg potential versus $\phi$.}
\label{Vcw2figure}
\end{center}
\end{figure}

As discussed in \cite{Luminogenesis2}, the inflaton is expected to decay to dark matter during entropy generation after inflation.  Using the slow-roll approximation, we examine 
a Coleman-Weinberg potential for inflation used in \cite{HMZ}:
\begin{equation}
V(\phi) = A (\phi + v)^4 \left[ \ln \frac{(\phi + v)^2}{v^2} - \frac{1}{2} \right] + \frac{A v^4}{2}.
\label{CW2}
\end{equation}
As in \cite{HMZ}, $\phi$ can be thought of as the physical field that is the real part of a scalar field $\Phi$ such that $\Phi^{\dag} \Phi = (\phi+v)^2$.  The 
expectation value of $\phi$ is $<\phi>=0$, but $<\Phi>=v$.  So the potential in terms of the full field $\Phi$ is this Coleman-Weinberg potential with 
$(\phi+v)^2 \rightarrow \Phi^{\dag}\Phi$, and it has its local maximum centered around $\Phi=0$, whereas 
the potential for $\phi$ has its local maximum centered around $\phi= -v$ (see Figure \ref{Vcw2figure}).  
So we use the potential in terms of $\phi$, which is mathematically equivalent up to a field shift 
of $v$, so the potential still displays the dynamics of a Coleman-Weinberg potential.  

The phase transition into the true vacuum energy with $<\Phi>=v$ will provide the symmetry breaking needed for the breaking of 
$SU(6) \rightarrow SU(4)_{DM} \times SU(2)_{L} \times U(1)_{DM}$.  So we expect $\mu_{DUT} = v$, and we constrain $v$ using inflation constraints from 
Planck.  

Two significant constraints from Planck \cite{Planck, Plancki}, assuming no running of the scalar spectral index and no tensor perturbations, come 
from the scalar spectral index, $n_s(k_{\star}) = 0.9603 \pm 0.0073$, and the scalar 
power-spectrum amplitude $A_s(k_{\star})$, $\ln(10^{10} A_s(k_{\star})) = 3.089_{-0.027}^{+0.024}$.  These values come from temperature power 
spectrum data from Planck and WMAP polarization at low multipoles and are obtained at the pivot scale $k_{\star}=0.05 ~{\rm Mpc}^{-1}$, and 
the scalar power spectrum with no running is modeled as $\mathcal{P}_{s}(k) = A_s(k_{\star}) \left( \frac{k}{k_{\star}} \right)^{n_s - 1}$.  
When tensor perturbations are considered with no running of the scalar or tensor spectral indexes, 
Planck reports at the 95\% confidence level $r(k_{\star}) \equiv \frac{A_s(k_{\star})}{A_t(k_{\star})} < 0.12$ for the tensor-scalar ratio and $n_s(k_{\star}) = 0.9624 \pm 0.0075$ using the 
same data sources mentioned earlier and obtained
at the pivot scale $k_{\star}=0.002 ~{\rm Mpc}^{-1}$.  We note that the latest joint analysis of the Planck and BICEP2/Keck data \cite{joint} 
overturn BICEP2's high estimate of $r$, which we do not consider in our 
analysis.  Although Planck also gives constraints on slow-roll parameters of higher order when running is allowed, 
depending on the data set used, they are consistent with no running and are not very helpful in constraining our model, so we do not present analysis concerning these 
constraints from Planck here.  

We use the slow-roll approximation to first order to apply these constraints, and we evaluate at the pivot scale's Hubble radius crossing, which is denoted by $\star$ 
as a superscript or subscript.
Natural units are used throughout with the reduced Planck mass $M_{Pl} \equiv (8 \pi G)^{-1/2}$. 

The scalar spectral index at the pivot scale in terms of the slow-roll potential parameters is given by 
\begin{equation}
n_s(k_{\star}) = 1 + 2 \eta_V^{\star} - 6 \epsilon_V^{\star},
\label{index}
\end{equation}
where 
\begin{equation}
\epsilon_V = \frac{M_{Pl}^2 }{2} \left(\frac{V_{,\phi}}{V} \right)^2
\label{eta}
\end{equation}
and 
\begin{equation}
\eta_V = M_{Pl}^2 \frac{V_{,\phi \phi}}{V}.
\label{epsilon}
\end{equation}
The scalar power spectrum amplitude is given by
\begin{equation}
A_s(k_{\star}) = \frac{1}{24 \pi^2} \frac{V_{\star}}{M_{Pl}^4} \frac{1}{\epsilon_V^{\star}},
\label{Ps}
\end{equation}
The tensor-scalar ratio at the pivot scale is 
\begin{equation}
r(k_{\star}) = 16 \epsilon_V^{\star},
\label{r}
\end{equation}
and the number of e-folds before the end of inflation when the pivot scale $k_{\star}$ exited the Hubble radius, $N_{\star}$, is given by
\begin{equation}
N_{\star} = \frac{1}{M_{Pl}^2} \int_{\phi_{end}}^{\phi_{\star}} \frac{V}{V_{,\phi}}.
\label{efolds}
\end{equation}

Plugging in our potential from Equation (\ref{CW2}) into Equation (\ref{index}), we obtain the relation 
\begin{equation}
\begin{split}
&v^2 ~ (n_s(k_{\star})-1)[x_{\star}(4+6x_{\star}+4x_{\star}^2+x_{\star}^3)-2(1+x_{\star})^4 \ln(1+x_{\star})^2]^2 = \\
&16 M_{Pl}^2 ~ (1+x_{\star})^2 [(4+4x_{\star}+6x_{\star}^2+4x_{\star}^3+x_{\star}^4) \ln (1+x_{\star})^2 - \\
&6(1+x_{\star})^4 [\ln (1+x_{\star})^2]^2 -2x_{\star} (4+6x_{\star}+4x_{\star}^2+x_{\star}^3)],
\end{split}
\label{vf}
\end{equation}
where $x \equiv \frac{\phi}{v}$.  In our potential in Figure \ref{Vcw2figure}, the domain of inflation begins soon after (to the right of)
 the local maximum at $\phi = - v$ (or $x=-1$), and we determine $x_{end}$ from the condition $\epsilon_{V}=1$, which is when 
 the acceleration of the universe due to inflation ceases, and this happens before entropy generation around the local minimum at the origin.  

Using Equations (\ref{index}) and (\ref{Ps}), we get
\begin{equation}
V_{\star}^{3} = 192 ~M_{Pl}^6 ~\pi^2 ~A_s(k_{\star}) ~(1+x_{\star})^6 ~[\ln(1+x_{\star})^2]^2 ~v^6.
\label{Vcw2}
\end{equation}

Evaluating the definite integral in Equation (\ref{efolds}), we get 
\begin{equation}
\begin{split}
&N_{\star} = - \frac{v^2}{16 M_{Pl}^2} \{ 2(x_{\star}-x_{end})(x_{\star}+x_{end}-34)+{\rm Ei}[-\ln (1+x_{\star})^2 ] \\
&- {\rm Ei}[-\ln (1+x_{end})^2 ] - {\rm li}[(1+x_{\star})^2] +{\rm li}[(1+x_{end})^2]   \},
\end{split}
\label{N}
\end{equation}
where ${\rm li} (y) \equiv \int_0^y \frac{ds}{\ln s}$ and ${\rm Ei} (y) \equiv \int_{-y}^{\infty} \frac{e^{-s}}{s} ds$.

Comparing Equations (\ref{vf}) and (\ref{Vcw2}) with the aforementioned constraints from Planck on $n_s$ and $A_s$, $v=\mu_{DUT}$ is given below for reasonable choices 
of the energy scale of inflation, $V_{\star}^{1/4}$.  
\begin{eqnarray}
V_{\star}^{1/4} = 10^{13} ~{\rm GeV}, ~A = (8.837 \pm 1.563)~10^{-6}, ~v = (2.192 \pm 0.098)~10^{14} ~{\rm GeV}, ~N_{\star}=79.25\pm14.57  \nonumber \\
V_{\star}^{1/4} = 10^{14} ~{\rm GeV}, ~A = (9.422 \pm 1.669)~10^{-6}, ~v = (2.157 \pm 0.096)~10^{15} ~{\rm GeV}, ~N_{\star}=79.49\pm14.61 \nonumber \\
V_{\star}^{1/4} = 10^{15} ~{\rm GeV}, ~A = (1.021 \pm 0.181)~10^{-5}, ~v = (2.114 \pm 0.095)~10^{16} ~{\rm GeV}, ~N_{\star}=79.88\pm14.68 \nonumber \\ 
V_{\star}^{1/4} = 10^{16} ~{\rm GeV}, ~A = (1.136 \pm 0.202)~10^{-5}, ~v = (2.059 \pm 0.093)~10^{17} ~{\rm GeV}, ~N_{\star}=80.63\pm14.80 \nonumber \\
V_{\star}^{1/4} = 10^{17} ~{\rm GeV}, ~A = (1.336 \pm 0.240)~10^{-5}, ~v = (1.977 \pm 0.090)~10^{18} ~{\rm GeV}, ~N_{\star}=83.80\pm15.26 \nonumber \\
\label{results} 
\end{eqnarray}
Using these results and Equation (\ref{r}), we find that $r \lesssim 10^{-8}$ for all these scenarios, so Planck's constraint on $r$ is satisfied.  The precise value of $N_{\star}$ 
depends on the energy scale of inflation and the uncertain details of entropy generation at the end of inflation (see, for example, Equation (24) of \cite{Plancki} for more details),
 but the value of $N_{\star}$ 
for each scenario above is greater than the minimum number required to solve the horizon problem for each energy scale.  
\begin{table}[h]
 \begin{center}
 \begin{tabular}{| l | l | l |} \hline
 Scenario & DM Mass for $\mu_{DUT}=10^{18}$ GeV & DM Mass for $\mu_{DUT}=10^{14}$ GeV \\ \hline
 $M_1 \sim \mu_{DUT}$ &  $4.7 \times 10^{12}$ & $2.9 \times 10^8$ \\
$M_1=10^{17} ~{\rm GeV}$ & $1.1 \times 10^{12}$ & - - \\
 $M_1=10^{16} ~{\rm GeV}$ & $2.6 \times 10^{11}$ & - - \\
 $M_1=10^{15} ~{\rm GeV}$ & $6.0 \times 10^{10}$ & - -\\
 $M_1=10^{14} ~{\rm GeV}$ & $1.4 \times 10^{10}$ & - -\\
$M_1=10^{13} ~{\rm GeV}$ & $3.3 \times 10^9$ & $6.8 \times 10^7$ \\
 $M_1=10^{12} ~{\rm GeV}$ & $ 7.6 \times 10^8$ & $1.6 \times 10^7$\\
 $M_1=10^{11} ~{\rm GeV}$ &  $ 1.8 \times 10^8$ & $3.7 \times 10^6$\\
 $M_1=10^{10} ~{\rm GeV}$ & $4.2 \times 10^7$ & $8.6 \times 10^5$\\
   $M_1=10^{9} ~{\rm GeV}$ & $9.7 \times 10^6$ & $2.0 \times 10^5$\\
  $M_1=10^{8} ~{\rm GeV}$ & $ 2.3 \times 10^6$ & $4.7 \times 10^4$\\
 $M_1=10^{7} ~{\rm GeV}$ & $ 5.3 \times 10^5$ & $1.1 \times 10^4$\\
  $M_1=10^{6} ~{\rm GeV}$ & $ 1.2 \times 10^5$ & $2.6 \times 10^3$\\
  $M_1=10^{5} ~{\rm GeV}$ & $2.9 \times 10^4$ & $6.0 \times 10^2$\\
   $M_1=10^{4} ~{\rm GeV}$ & $6.8 \times 10^3$ & $1.4 \times 10^2$\\ \hline
 \end{tabular}
 \end{center}
\caption{\label{table3} Upper and lower bounds on dark matter dynamical mass derived from a range of $\mu_{DUT}$ from Equation (\ref{results}) for various scenarios (with $M_2 \sim \mu_{DUT}$).  Values are only provided for scenarios in which $M_1 \lesssim \mu_{DUT}$.  All values are given in GeV.  The small value of the dark matter dynamical mass lower bound on the last row is given 
without concern for the phenomenological constraints on mirror-fermion masses, which are discussed in \cite{PQnew}.}
 \end{table}

\begin{figure}
\begin{center}
\fbox{\includegraphics[scale=1.6]{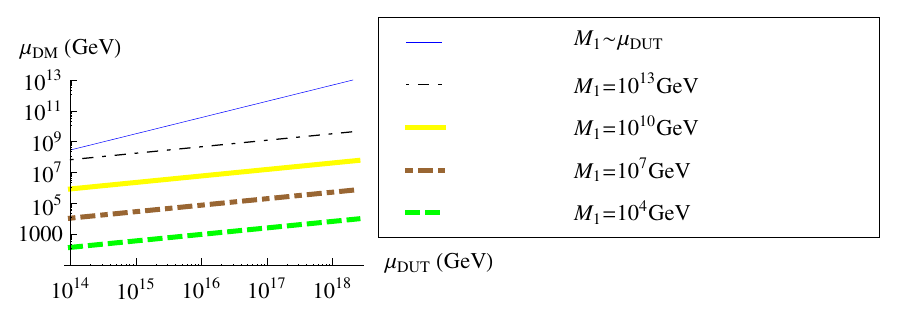}}
\caption{ $\mu_{DM}$ is shown for a few chosen scenarios (with $M_2 \sim \mu_{DUT}$).  
}
\label{DMmassplotfig}
\end{center}
\end{figure}

The constrained values of $\mu_{DUT}$ in Equation (\ref{results}) imply values for dark matter dynamical mass depending on what 
$M_1$ and $M_2$ are.  The values of $v=\mu_{DUT}$ in Equation (\ref{results}) range from $10^{14}$-$10^{18}$ GeV.  In Table \ref{table3}, using Equation (\ref{muDM}) for various scenarios concerning the choice of $M_1$, we give the upper and lower bounds for dark matter dynamical mass based on this range for $\mu_{DUT}$. 
A wide range of predictions of dark matter dynamical mass
is exemplified in Figure \ref{DMmassplotfig} for various scenarios.   
There are no strong astrophysical constraints on the scale of $M_1$, and the number density of these particles is in general below a detectable level 
due to Boltzmann suppression.  
In fact, we even show scenarios in which $M_1$ is as low as $10^{4} ~{\rm GeV}$ since the 
Boltzmann suppression factor ($e^{-M/T}$) in the number density for particles of mass $10^4 ~{\rm GeV}$ in equilibrium would be 
$e^{-100}$ and stronger for temperatures of $T \sim 100~{\rm GeV}$ (electroweak scale) and lower.  Since a candidate for dark matter with moderately low mass 
is in general desirable, we show that such a dynamical mass is attainable with the appropriate choice of $M_1$.  If or when the dark matter mass is more precisely 
constrained in the future, we can then know what the scale of $M_1$ is.  $M_1$ particles could be produced in high-energy astrophysical 
phenomena such as active galactic nuclei or quasars, and they would annihilate into radiation detected on Earth.  We could constrain the cross section of such particles 
once the scale of $M_1$ is deduced from the mass of dark matter.  Note that the inflaton, with mass 
$m_{\phi}= \sqrt{8 A} v$, decays to two dark matter
particles, which are massless until the confinement scale $\mu_{DM} \ll \mu_{DUT}$.

\bigskip
\begin{center}
{\bf Conclusion}
\end{center}

In this paper, we explore the unique connection in the luminogenesis model, a model that consistently combines dark matter and the Standard Model, 
between cosmic inflation and the creation of dark matter that allows 
the constraining of the dynamical mass of dark matter particles.  The constraint on $\mu_{DUT}$ is obtained by fitting a particular symmetry-breaking potential, although the connection between the unification scale $\mu_{DUT}$ and the dark matter dynamical 
mass $\mu_{DM}$ remains independent of the supposed correct model of inflation.  

Since the descent into the true vacuum of inflation triggers the breaking of the DUT symmetry and thus the conversion of the inflaton to dark matter, 
we can arrive at rough constraints on the DUT scale via constraints on inflation from cosmological probes.  Through picking a Coleman-Weinberg potential and 
reasonable energy scales for inflation, we 
arrive at constraints on the DUT scale, and we can then derive an upper bound on the dark matter dynamical mass via RG flow.  
We run the $SU(2)_L$ coupling from the known electroweak scale 
up to the DUT scale, where it is unified with $SU(4)_{DM}$.  We then run the $SU(4)_{DM}$ coupling down to its confinement scale, taken to be when $\alpha_4 \sim 1$, 
and we arrive at the dynamical mass scale for dark matter.  Various dynamical mass values for dark matter are possible depending on the mass scale $M_1$ and $M_2$.  The possibility 
of strongly self-interacting dark matter, as proposed in \cite{Luminogenesis1, Luminogenesis2}, with dynamical masses obtained through the connection $\mu_{DUT} \rightarrow \mu_{DM}$ as studied here has wide 
implications concerning the resolution of dwarf galaxy structures and dark matter cusps at the centers of galaxies and its potential detectability.  This is under investigation.


\bigskip

\begin{center}

{\bf Acknowledgements}

\end{center}

PQH is supported in part by the U.S. Department of Energy under
Grant No. DE-FG02-97ER41027.  KJL is supported by the Pirrung Foundation.

\bigskip

\end{document}